\documentclass[
showkeys,11pt,
preprint,preprintnumbers,nofootinbib,
groupedaddress,superscriptaddress,amsmath,amssymb]{revtex4}
\usepackage{graphicx}
\usepackage{dcolumn}
\usepackage{bm}
\usepackage{amssymb}
\usepackage{amsmath}
\usepackage{epsfig}    
\usepackage{color}
\usepackage{slashed}
\usepackage{hhline}

\def\be{\begin{equation}}
\def\ee{\end{equation}}
\newcommand{\bea}{\begin{eqnarray}}
\newcommand{\eea}{\end{eqnarray}}
\newcommand{\nn}{\nonumber}

\numberwithin{equation}{section}

\begin{document}

{\begin{flushright}{KIAS-P16008}
\end{flushright}}

\title{A four-loop Radiative Seesaw Model}
%
\author{Takaaki Nomura}
\email{nomura@kias.re.kr}
\affiliation{School of Physics, KIAS, Seoul 130-722, Korea}

\author{Hiroshi Okada}
\email{macokada3hiroshi@cts.nthu.edu.tw}
\affiliation{Physics Division, National Center for Theoretical Sciences, Hsinchu, Taiwan 300}

\date{\today}

\begin{abstract}
We propose a new type of radiative neutrino model with global hidden $U(1)$ symmetry, in which neutrino masses are induced at the four loop level.
Then we discuss the muon anomalous magnetic moment to solve the discrepancy between observation and the standard model prediction, 
and estimate the relic density of a fermion or boson DM candidate in the model.
We also discuss the diphoton resonance $R$ by considering a process $pp \to R \to \gamma \gamma$ as a possible signal of our model. 
\end{abstract}
\maketitle
\newpage

\section{Introduction}

A radiative seesaw model is one of the attractive scenario to generate active neutrino masses. In such a model, some exotic particles with nonzero electric charges (bosons or fermions) are introduced in order to explain the tiny neutrino masses such as Zee-Babu model~\cite{zee-babu}.  Moreover there often exist dark matter (DM) candidates, which also  play a role in generating neutrino masses.
These exotic particles can induce interesting effects which would be observed in experiments such as the Large Hadron Collider (LHC). 

In 2015, a hint of new particle ($\Phi_{\rm New}$) is indicated by the ATLAS and CMS 
by the observation of the diphoton invariant mass spectrum at the LHC 13 TeV~\cite{ATLAS-CONF-2015-081,CMS:2015dxe,Aaboud:2016tru,Khachatryan:2016hje}.
The excess of the events could be interpreted as a production of $\Phi_{\rm New}$ decaying into two photons where
a vast of paper along this line of issue has been recently arisen in Ref.~\cite{Kobakhidze:2015ldh,Chao:2015nac,Kanemura:2015bli,Nomura:2016fzs,Modak:2016ung,Dutta:2016jqn,Deppisch:2016scs,Borah:2016uoi,Hati:2016thk,Yu:2016lof,Ding:2016ldt}
which are related to neutrino mass model, in order to give reasonable explanations or interpretations.
However new LHC data for the diphoton signal is announced which disfavor the excess~\cite{CMS:2016crm, ATLAS:2016eeo}. 
Although the diphoton excess is more like a statistical fluctuations, the studies on this issue indicate that 
the diphoton signal can be a good probe of a scalar (or pseudo scalar) boson which couples to fields with color and/or electric charge. 

In our paper, we propose a new type of radiative neutrino model with global hidden $U(1)$ symmetry, in which neutrino masses are induced at the four loop level. 
In our model, several charged fermions and bosons are introduced to provide a four loop diagram for generating neutrino mass matrix. 
Furthermore, in our setup, there exist dark matter (DM) candidates which are fermion or boson. 
We also explain the discrepancy of the muon anomalous magnetic moment to the standard model (SM) by using the exotic charged fermions.
Then we discuss diphoton resonance as a possible signal of our setup by considering two Higgs doublet sector which is the same as type-II two Higgs doublet model (2HDM).
For diphoton resonance, we focus on the CP-even neutral scalar boson since CP-odd scalar cannot have a trilinear coupling to the charged bosons. 
Indeed we take into account the consistency with observed SM Higgs properties such as branching ratio, since the CP-even scalar must influence to these observables.

In Sec.~II, we introduce our model and derive some formulas including neutrino mass matrix, muon anomalous magnetic moment, and the relic density of DM (fermion and boson case).
In Sec.~III, we discuss the diphoton resonance in our model. In Sec.~IV, we have  numerical analyses.
We conclude and discuss in Sec.~V.

\section{ Model setup and Analysis}

In this section, we explain our model with a hidden $U(1)$ symmetry. 
We also derive the formulas for neutrino mass matrix, muon $g-2$ and relic density of dark matter.

\subsection{Model setup}
 \begin{widetext}
\begin{center} 
\begin{table}[tb]
\begin{tabular}{|c||c|c|c|c|c||c|c|c|c|c|c|}\hline\hline  
&\multicolumn{5}{c||}{Lepton Fields} & \multicolumn{6}{c|}{Scalar Fields} \\\hline
& ~$L_L$~ & ~$e_R^{}$~ & ~$L'^{}_{}$ ~ & ~$E$ ~ & ~$N_R$~  & ~$\Phi_{1,2}$~ & ~$S$  & ~$S^{a +}$  & ~$k_1^{a+}$  & ~$k_2^{a+}$  & ~$\varphi$ \\\hline 
$SU(2)_L$ & $\bm{2}$  & $\bm{1}$  & $\bm{2}$ & $\bm{1}$ & $\bm{1}$ & $\bm{2}$ & $\bm{1}$  & $\bm{1}$   & $\bm{1}$   & $\bm{1}$   & $\bm{1}$ \\\hline 
$U(1)_Y$ & $-1/2$ & $-1$  & $-3/2$ & $-1$ & $0$ & $1/2$ & $0$ & $1$  & $1$  & $1$  & $0$  \\\hline
$U(1)$ & $0$ & $0$  & $n$ & ${n}$ & ${n}$ & $0$ & ${n}$ & $-n$  & $0$ & $0$  & $2n$  \\\hline
$Z_2$ & $+$ & $+$  & $+$ & ${+}$ & ${-}$ & $+$ & ${+}$ & $+$  & $-$ & $+$  & $+$  \\\hline
\end{tabular}
\caption{Contents of fermion and scalar fields
and their charge assignments under $SU(2)_L\times U(1)_Y\times U(1)$. We introduce $N_B$ sets of charged scalar fields, i.e. $a= 1,...,N_B$.}
\label{tab:1}
\end{table}
\end{center}
\end{widetext}

The particle contents and their charges are shown in Tab.~\ref{tab:1}.
We add vector-like exotic SU(2) doublet charged fermion $L'$ with $-3/2$ hypercharge and singlet $E'$ with $-1$ hypercharge, Majorana fermions $N_R$,
$N_B$ sets of three singly charged scalars $S^{a \pm}$, and $k^{a \pm}_{1(2)}$ ($a=1,...,N_B$) with different quantum numbers, two neutral scalars $S$ and $\varphi$, and one additional Higgs doublet to the SM.
{Here we emphasize that types of these exotic field contents are minimal combination to realize our new type of four loop diagram for neutrino mass generation which is shown below 
while forbidding neutrino mass generation at lower loop level. We then assumed multiplicity of $SU(2)$ singlet charged scalars to investigate enhancement effect in both neutrino mass generation and diphoton decay rate of heavy neutral scalar boson. }
In our model we require that only the two Higgs doublets $\Phi_{i}$  and $\varphi$ have vacuum expectation values (VEVs), which are respectively symbolized by $v_i/\sqrt2$ and $v'/\sqrt2$. 
The quantum number $n\neq0$ under the hidden $U(1)$ symmetry is arbitrary, but its assignment for each field is unique to realize our four loop neutrino model. 
After global U(1) breaking, we also have another $\tilde Z_2$ symmetry where particles with charge $n$ are odd under the symmetry.
Then we have a stable particle when it is the lightest one with odd parity under the $Z_2$ or $\tilde Z_2$ symmetries, which can be a DM candidate if it is neutral. 
Therefore our DM candidates are the lightest Majorana fermion $N_R|_{\rm lightest} = X$ and/or the lightest isospin singlet scalar $S\equiv (S_R+i S_I)/\sqrt2$.
Here we identify the first generation of $N_R$ or $S_I$ as a dark matter candidate respectively.
We also introduce an additional softly-broken $Z_2'$ symmetry where second Higgs doublet $\Phi_2$ and right-handed up-type quarks are assigned to parity odd under this symmetry. 
Thus the Yukawa coupling for two Higgs doublets with SM fermions is that of Type-II 2HDM.

The relevant Lagrangian and Higgs potential under these symmetries are given by
\begin{align}
-\mathcal{L}_{Y}
&\supset { M_{L}} \bar L'_{} L'_{}  + { M_{E}} \bar E_{} E_{} + (
y_{\ell} \bar L_{L} \Phi_1 e_{R}+
f_{}^a \bar L_{L} L'_{R} S^{a+}
+g_{}^a \bar E_{L} N_R  k_1^{a -}  \nn \\
& + y_R \bar L'_{L} \tilde \Phi_1 E_R  
{+ y_L \bar L'_{R} \tilde \Phi_1 E_L}
+ y_S \bar e_R E_L S^*  + \frac{y_N }{2} \varphi^* \bar N^c_R N_R + h.c.)  \label{Eq:lag-flavor}   
\end{align}
\begin{align}
V =& \ m_S^2 |S|^2 + m_\varphi^2 |\varphi|^2 + m_{k_1^a}^2 k_1^{a+}k_1^{a-} + m_{k_2^a}^2 k_2^{a+}k_2^{a-} + m_{S^+}^2 S^+ S^- \nonumber \\
& + \lambda_S |S|^4 + \lambda_\varphi |\varphi|^4 + \mu_{ab} (S S^{a+} k_2^{b-} +h.c.) +  \frac{\mu'}2 (\varphi^* S^2 +h.c.)
+ \lambda_{0}^{abcd} (k_1^{a +} k_2^{b -})(k_1^{c +} k_2^{d -})   \nonumber \\
& + m_1^2 |\Phi_1|^2 + m_2^2 |\Phi_2|^2 - m_3^2 (\Phi_1^\dagger \Phi_2 + h.c.) \nonumber \\
 &+ \frac{1}2{\lambda_1} |\Phi_1|^4 + \frac{1}{2} \lambda_2 |\Phi_2|^4 + \lambda_3 |\Phi_1|^2 |\Phi_2|^2 + \lambda_4 |\Phi_1^\dagger \Phi_2|^2 + \frac{1}{2} \lambda_5 [(\Phi^\dagger_1 \Phi_2)^2 + h.c.] \nonumber \\
 & + \lambda_{\Phi_1 k_{1}}^{ab} (\Phi_1^\dagger \Phi_1)(k_{1}^{a+} k_{1}^{b-}) + \lambda_{\Phi_2 k_{1}}^{ab} (\Phi_2^\dagger \Phi_2)(k_{1}^{a+} k_{1}^{b-}) \nonumber \\
  & + \lambda_{\Phi_1 k_{2}}^{ab} (\Phi_1^\dagger \Phi_1)(k_{2}^{a+} k_{2}^{b-}) + \lambda_{\Phi_2 k_{2}}^{ab} (\Phi_2^\dagger \Phi_2)(k_{2}^{a+} k_{2}^{b-}) \nonumber \\
& + \lambda_{\Phi_1 S^+}^{ab} (\Phi_1^\dagger \Phi_1)(S^{a+} S^{b-}) + \lambda_{\Phi_2 S^+}^{ab} (\Phi_2^\dagger \Phi_2)(S^{a+} S^{b-}) \nonumber \\
& + \lambda_{\Phi_1 S} (\Phi_1^\dagger \Phi_1)|S|^2 + \lambda_{\Phi_2 S} (\Phi_2^\dagger \Phi_2)|S|^2
+ \lambda_{\Phi_1 \varphi} (\Phi_1^\dagger \Phi_1)|\varphi|^2 + \lambda_{\Phi_2 \varphi} (\Phi_2^\dagger \Phi_2)|\varphi|^2+ \cdots \label{eq:potential}
, 
\end{align}
where the flavor indices are abbreviated for brevity, and we omitted some quartic terms containing only $\{ \varphi, S, k_1^+, k_2^+, S^+ \}$ which are irrelevant in our analysis.
After the global $U(1)$ spontaneous breaking by $\langle \varphi \rangle =v'/\sqrt{2}$, we obtain the Majorana masses  $M_N\equiv y_N v'/\sqrt2$.
The first term of $\mathcal{L}_{Y}$ generates the SM
charged-lepton masses $m_\ell\equiv y_\ell v_1/\sqrt2$ after the spontaneous breaking of electroweak symmetry by $\langle \Phi_i \rangle = v_i/\sqrt{2}$.
We work on the basis where all the massless coefficients are real and positive for simplicity. 
In our analysis, we assume $\lambda_{\Phi_i \varphi}$ is negligibly small so that mixing between $\varphi$ and neutral components of the doublets  are ignored.
Then VEVs and masses of Higgs doublets are obtained same as Type-II 2HDM.
The isospin doublet scalar fields can be parameterized as $\Phi_i=[w_i^+,\frac{v_i+h_i+iz}{\sqrt2}]^T$
where $v~\simeq 246$ GeV is VEV of the Higgs doublet, and one component of $w_i^\pm$
and $z_i$ are respectively absorbed by the longitudinal component of $W$ and $Z$ boson.
The isospin singlet scalar field can be parameterized by $\varphi=\frac{v'+\phi}{\sqrt2}e^{2i n G/v'}$ where mixing between $\phi$ and $h_i$ is negligible in our assumption 
 and the G is a Goldstone boson associated with symmetry breaking of the global U(1).
Then we focus on the CP-even Higgs where mass eigenstates are 
\begin{equation}
\begin{pmatrix} h_1 \\ h_2 \end{pmatrix} = \begin{pmatrix} \cos \alpha & - \sin \alpha \\ \sin \alpha & \cos \alpha \end{pmatrix} \begin{pmatrix} H \\ h \end{pmatrix}
\end{equation}
where $h$ and $H$ denote SM Higgs and heavier CP-even Higgs respectively.
In our analysis of $H$ production via gluon fusion, we focus on the Yukawa interactions of $H$ and top quark  
\begin{align}
{\cal L}^Y \supset &  -   \frac{m_{t}}{v} \frac{\sin \alpha}{\sin \beta} \bar t t H,
\end{align}
where $\tan \beta = v_2/v_1$ as usual.
The $h W^+W^-/ZZ$ and $H W^+W^-/ZZ$ couplings are respectively proportional to $\sin (\alpha - \beta)$ and  $\cos (\alpha - \beta)$ in the 2HDM~\cite{Gunion:1989we}. 
In this paper, we assume alignment limit~\cite{Carena:2013ooa}, $\alpha - \beta = \pi/2$, to suppress $H \to W^+W^-/ZZ$ decay channel.

 It is worth mentioning some issues on the  goldstone boson $G$ that could plays significant roles in particle physics and cosmology.
The first issue is that an effect on cosmic microwave background via cosmic string generated by the spontaneous breaking of the global $U(1)$ symmetry.
It  possibly puts a constraint on our scenario. The bound discussed in ref.~\cite{Battye:2010xz} can be interpreted as $v' \le 10^{15}$ GeV {which can be easily satisfied since VEVs are O(100)-O(1000) GeV in our model}. 
The second one is that $G$ would induce a discrepancy of the effective number of neutrino species in the early Universe, which is denoted by $\Delta N_{\rm eff}$. The recent data reported by Planck shows $\Delta N_{\rm eff}=0.04\pm0.33$  at the 95 \% confidential level~\cite{Ade:2015xua}. In our case, $\Delta N_{\rm eff}$ is about 0.052. Therefore, our model can evade this constraint. 
Moreover we do not have a tree level interaction which provides a force mediated by the goldstone boson so that further constraint will not be imposed.

{\it Exotic Charged Fermion mass matrix}:
The singly exotic charged fermion mass matrix is given by 
\begin{align}
{\cal L}_{\text{mass}} &= -(\overline{E^{-}},\overline{e'^{-}})
\begin{pmatrix}
M_E &   m'_{} \\
m'_{} & M_L
\end{pmatrix}
\begin{pmatrix}
E^{-} \\
e'^{-}
\end{pmatrix} + \text{h.c.}=
-(\overline{E_1^{}},\overline{E_2^{}})
\begin{pmatrix}
M_{E^1} &   0 \\
0  & M_{E^2}
\end{pmatrix}
\begin{pmatrix}
E_{1}^{} \\
E_{2}^{}
\end{pmatrix} + \text{h.c.} , 
\end{align}
where we define $L'\equiv [e'^{-},e'^{--}]^T$, $m'_{} =  \frac{v}{\sqrt{2}}  y_R $ {assuming $ y_L = y_L^T = y_R= y_R^T $ for simplicity.} 
The mass eigenstates $E^1$ and $E^2$ are defined by the bi-unitary transformation:
\begin{align}
\begin{pmatrix}
E^{-} \\
e'^{-}
\end{pmatrix}
=
\begin{pmatrix}
c_{\theta_E} & -s_{\theta_E} \\
s_{\theta_E} &c_{\theta_E}
\end{pmatrix}
\begin{pmatrix}
E_1^{-} \\
E_2^{-}
\end{pmatrix}, 
\end{align}
where $s_{\theta_E} \equiv \sin\theta_E$ and $c_{\theta_E} \equiv \cos\theta_E$.
The mass eigenvalues and the mixing angles $\theta_E$ are respectively given by 
\begin{align}
M_{E^{1,2}} &= \frac{1}{2}\left(M_E + M_L \mp\sqrt{(M_E - M_L)^2 + 4m'^2}\right), \quad
\tan2\theta  = \frac{2 m'}{M_{E} - M_L}. 
\end{align}
Notice here that the mass of the doubly charged fermion $e'^{\pm\pm}$ is given by $M_L$.
We also note that large mass splitting in components of $SU(2)$ doublet $L'$ due to $m' = v y_R/\sqrt{2}$ 
would provide sizable contribution to $T$-parameter. Then we consider similar size of mass for $M_L$ and $M_{E^{1,2}}$ 
in our numerical analysis below.

%
\begin{figure}[tb]
\begin{center}
\includegraphics[width=80mm]{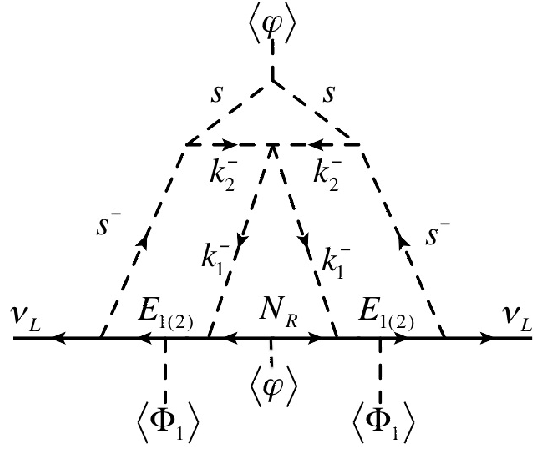}
\caption{ Neutrino masses at the four-loop level. 
{The arrows in the diagrams indicate chirality flow for neutral fermion lines, electric charge flow for boson lines and the both flows for charged fermion lines.
For lines with $k_{1,2}^-$ and $S^-$, $N_B$ number of charged boson can propagate. At the top of diagram we have $v' \mu'$ factor from $\mu' \langle \varphi* \rangle S^2/2$ term in the Lagrangian.  
}
}   \label{fig:neut1}
\end{center}\end{figure}
\subsection{ Neutrino mass matrix}
%
The leading contribution to the active neutrino masses $m_\nu$  is given at four-loop level as shown in Figure~\ref{fig:neut1}, and its formula is given as follows:
\begin{align}
(m_{\nu})_{ij}&\approx \frac{2 N_B^6 \lambda_0 \mu^2(m^2_{R}-m^2_I) s_{\theta_E}^2  c_{\theta_E}^2 }{(4\pi)^8 M^6_{\rm max} } 
\sum_{\alpha,\alpha'=1}^2\sum_{k,\ell,m=1}^3 (f_{ik} M_{E^\alpha_k} g_{k\ell} M_{N_\ell} g_{m\ell} M_{E^{\alpha'}_m} f_{jm})G (x_I),\nn\\
&\approx \frac{N_B^6 \lambda_0 v' \mu' \mu^2  s_{2\theta_E}^2  }{4\sqrt2(4\pi)^8 M^6_{\rm max} } 
\sum_{\alpha,\alpha'=1}^2\sum_{k,\ell,m=1}^3 (f_{ik} M_{E^\alpha_k} g_{k\ell} M_{N_\ell} g_{m\ell} M_{E^{\alpha'}_m} f_{jm})G (x_I),
\label{eq:neutrinoM}
\end{align}
\begin{align}
G (x_I)&\equiv G\left( \frac{m_{k_1^\pm}^2}{M^2_{\rm max}}, \frac{M_{E^{\alpha'}_m}^2}{M^2_{\rm max}},
\frac{m_{S^\pm}^2}{M^2_{\rm max}}, \frac{m_{k_2^\pm}^2}{M^2_{\rm max}}, 
 \frac{m_{S_R}^2}{M^2_{\rm max}}, \frac{m_{S_I}^2}{M^2_{\rm max}},\frac{M_{N_\ell}^2}{M^2_{\rm max}},
\frac{M_{E^\alpha_k}^2}{M^2_{\rm max}} \right)\nn\\
&=\int\Pi_{i=1}^6 dx_i \frac{\delta(\sum_{i=1}^6 x_i-1)}{(x_3^2-x_3)^2}\int dadbdc\frac{a^3\delta(a+b+c-1)}{D^4}
\int d\alpha d\beta d\gamma\frac{\alpha^3\delta(\alpha+\beta+\gamma-1)}{G^4}\nn\\
&\times
\int d\rho d\sigma \frac{\rho^3\delta(\rho+\sigma-1)}{(\rho \frac{H}{G}-\sigma X_{k_2})^3}
,
\end{align}
\begin{align}
H&= \frac{a \alpha(x_3 X_{k_1} + x_1 X_{E_m}+x_2 X_{S^\pm}+x_4 X_{k_2}+x_5 X_{S_R}+x_6 X_{S_I})}{(x_3^2-x_3)D}\nn\\
&-
\frac{\alpha(b X_{k_1} - c X_{N_\ell})}{D} +\beta X_{E_k} +\gamma X_{S^\pm}, \
D=\frac{a x_4(x_4+x_3-1)}{(x_3-1)^2-x_3}-b,
\nn\\
G&=\frac{\alpha^2}{D^2}\left(\frac{(x_3 x_5+x_3x_6+x_3+x_4-1)a x_4}{(x_3-1)^2 x_3}\right)^2\nn\\
&-
\frac{a\alpha}{x_3(x_3-1)^2D}\left[x_4 (x_4+x_3-1)+x_3 (x_5+x_6)(x_4+x_5+x_6)\right],
\label{mnu1}
\end{align}
where each of $m_{R}$ and $m_{I}$ is the mass of $S_R$ and $S_I$, and satisfies $m_R^2-m^2_I=\mu'v'/(2\sqrt2)$, 
and we assume the coupling constants are same for different charged scalar sets so that $N_B^6$ is multiplied. 
Here we define $M_{\rm max}={\rm Max}[M_E,M_N, m_{S^\pm},m_{k_{1/2}^\pm},m_{S_R},m_{S_I}]$, 
$m_\nu$ should be $0.001\ {\rm eV}\lesssim m_\nu \lesssim 0.1\ {\rm eV}$ from the neutrino oscillation data~\cite{pdf}.
Note that the loop diagram Fig.~\ref{fig:neut1} contains only exotic particles inside the loop.
Here the lepton number violation arises from the line of $N_R$ after the spontaneous breaking of the global U(1) by $\varphi$ as shown in Fig.~\ref{fig:neut1}. And our model induces the neutrino mass through the dimension 7 operator which is also found in Fig.~\ref{fig:neut1}, while  both of Zee and Zee-Babu model induce dimension 5 operator for neutrino mass generation.

\subsection{ Muon anomalous magnetic moment}

The muon anomalous magnetic moment (muon $g-2$) has been 
measured at Brookhaven National Laboratory
that {indicates} a discrepancy between the
experimental data and the prediction in the SM. 
The difference $\Delta a_{\mu}\equiv a^{\rm exp}_{\mu}-a^{\rm SM}_{\mu}$
is calculated in Ref.~\cite{discrepancy1} and Ref.~\cite{discrepancy2}, giving the values respectively as 
\begin{align}
\Delta a_{\mu}=(29.0 \pm 9.0)\times 10^{-10},\
\Delta a_{\mu}=(33.5 \pm 8.2)\times 10^{-10}. \label{dev12}
\end{align}
The above results 
correspond
to $3.2\sigma$ and $4.1\sigma$ deviations, respectively. 
In our model, contribution to $\Delta a_\mu$ is induced at the one-loop level where exotic fermions and bosons propagate inside loop diagrams. 
Calculating one-loop diagrams, formula of muon $g-2$ is given by
\begin{align}
&\Delta a_\mu\approx \frac{m_\mu^2}{(4\pi)^2}
\biggl[ N_B |f|^2_{22} \left[F(e'^{--},S^\pm) + 2F(S^\pm,e'^{--})\right] \nonumber \\
& \qquad \qquad \qquad \qquad +\frac{|y_S|_{22}}{2} \sum_{\alpha,\beta=R,I} 
\left[c^2_{\theta_E}   F(S_{\alpha},E_1^{-}) + s^2_{\theta_E}   F(S_{\alpha},E_2^{-})\right]
\biggr],\label{eq:g-2}
\\
&F(x,y)\approx \frac{
2 m_x^6 +3 m_x^4 m_y^2 - 6 m_x^2 m_y^4 +m_y^6 + 12 m_x^4 m_y^2 \ln\left[\frac{m_y^2}{m_x^2}\right]}
{12(m_x^2- m_y^2)^4}
\label{damu},
\end{align}
where we assume $f^a$ is same value for different charged scalar sets.

It is worth mentioning that the lepton flavor violating (LFV) processes are always {induced} by the same interactions generating the muon anomalous magnetic moment.
In our case, LFVs are generated from the terms proportional to $f$ and $y_S$ at the one-loop level, and  these couplings or masses related to exotic fermions or bosons are constrained. The stringent bound is given by the $\mu\to e\gamma $ process at the one loop level~\cite{Adam:2013mnn}, and its branching ratio is given by
\begin{align}
& {\rm BR}(\mu\to e\gamma)
\approx \frac{3 \alpha_{\rm em}}{64\pi {\rm G_F^2}}
\biggl| N_B |f|^2_{21} \left[F(e'^{--},S^\pm) + 2F(S^\pm,e'^{--})\right]  \\
& \qquad \qquad \qquad \qquad +\frac{|y_S|_{21}}{2} \sum_{\alpha,\beta=R,I} 
\left[c^2_{\theta_E}   F(S_{\alpha},E_1^{-}) + s^2_{\theta_E}   F(S_{\alpha},E_2^{-})\right]
\biggr|^2 \lesssim 5.7\times 10^{-13},\nn
\label{eq:mueg}
\end{align}
where $\alpha_{\rm em}\approx 1/128$ is the fine structure constant, and ${\rm G_F}\approx 1.17\times 10^{-5}$[GeV$^{-2}$] is the Fermi constant.
The upper bound of the off-diagonal Yukawa coupling squares can typically be estimated as $|f|^2_{21} \approx |y_S|^2_{21}\approx {\cal O}(10^{-4})$.
Here we fix the related values to be $s_{\theta_E}\approx1/\sqrt2$, $M_{E^{1/2}}=M_L=m_R=m_I\approx 500$ GeV, $m_{\phi^\pm}\approx 380$ GeV, and $N_B=10$ such that  ${\rm BR}(\mu\to e\gamma)$ becomes maximum within the range of the numerical analysis.
Thus once we assume $f$ and $y_S$ to be diagonal, such LFVs can  simply be evaded.
\footnote{
We also note that even if off-diagonal components appear at one-loop level, one can satisfy bounds of LFVs when
$|f|_{21}, |y_S|_{21} \lesssim 0.01 $ are taken.
}
Even in this case, 
the neutrino mixings are 
induced via the coupling of $g$. Hence we retain the consistency of the LFV constraints without conflict of the neutrino oscillation data and the muon anomalous magnetic moment, 
applying this assumption to the numerical analysis.

\subsection{ Dark matter}
{\it Case 1.  Fermion DM} : 
First of all, we assume the lightest component of Majorana particle $N_R$ is our DM candidate, which is denoted by $X$. 
Then we find that the dominant DM annihilation process is $2 X \to 2 G$ which can provide the observed relic density $\Omega h^2\approx0.12$~\cite{Ade:2013zuv}.
The non-relativistic cross section for $2 X\to  2 G$ in $s$-channel is given by 
\begin{align}
\sigma v_{\rm rel}\approx 
\frac{n^2 M^6_X s^2 (s-4 M_X^2)}{32\pi v'^4 [(s - m_\phi^2)^2+m_\phi^2 \Gamma_\phi^2 ]},
\end{align}
where $\Gamma_\phi$ is the decay rate of $\phi$ and its concrete formulae are found in ref.~\cite{Nishiwaki:2015iqa}. 
Notice here we neglect that the mixing between $h$ and $\phi$ so that $X$ does not interact with quark sector.
Therefore, the spin independent scattering cross section vanishes at the tree level. 
The measured relic density is obtained at around the pole of $M_X\approx m_\phi/2$.
In this case, $s$ should directly be integrated out from $s\approx4M^2_X$ to infinity.
Here we have followed the formula of refs.~\cite{Nishiwaki:2015iqa} and \cite{Gondolo:1990dk}  to get the relic density in our numerical analysis where it is approximately given by
\begin{align}
\Omega h^2\approx \frac{1.07\times10^9  [{\rm GeV}]^{-1}} 
{g^{1/2}_* M_{\rm pl} \int_{x_f}^{\infty} \frac{dx}{x^2} \langle \sigma v_{\rm rel} \rangle_{\rm anni}},
\label{eq:relic}
\end{align}
where $M_{\rm pl}=1.22\times 10^{19}[{\rm GeV}] $ is the Planck mass, $\langle \sigma v_{\rm rel} \rangle_{\rm anni}$ is thermal average of $\sigma v_{\rm rel}$ which is 
a function of  $x \equiv m_{DM}/T$ with temperature $T$, $x_f(\approx 25)$ is $x$ at the freeze out temperature and $g_* (\approx100)$ is the total number of effective relativistic degrees of freedom at the time of freeze-out.

\noindent
{\it Case 2.  Boson DM} : 
Next, we consider the bosonic DM, assuming $S_I$ as the DM candidate.
In this case, we find three DM annihilation processes to provide a cross section explaining the relic density.
The dominant annihilation processes are $2 X\to  \ell\bar\ell$ with $t,u$-channels,
\footnote{We neglect $t,u$-channels for simplicity.}
$2 X\to 2h$ with contact interaction, and  $2 X\to  2G$ with contact interaction and $t,u$-channels. 
The formulae of non-relativistic cross sections for these processes are respectively given by 
\begin{align}
\sigma v_{\rm rel} &\approx \sigma v_{\rm rel}(2X\to \ell\bar\ell) + \sigma v_{\rm rel}(2X\to 2h) +\sigma v_{\rm rel}(2X\to 2G) ,\\
\sigma v_{\rm rel}(2X\to \ell\bar\ell) &\approx 
\frac{|y_S|^4 M^6_X}{240\pi}
\left[\frac{c^4_{\theta_E}}{(M_{E^1}^2+ M^2_X)^4}+ \frac{s^4_{\theta_E}}{(M^2_{E^2}+ M^2_X)^4}\right] v^4_{\rm rel},\\
\sigma v_{\rm rel}(2X\to 2h) &\approx 
\frac{|\lambda_{h S}|^2}{64\pi M^2_X}\sqrt{1-\frac{m^2_h}{M_X^2}} ,\\
\sigma v_{\rm rel}(2X\to 2G) &\approx 
\frac{n^4 \mu'^2}{8 M_X^2}
\left[ \left(\frac2{v'} - \frac{2\mu'}{m_R^2+M_X^2}\right)^2 +\frac{4 m_R^2 M_X^2 \mu' (m_R^2+M_X^2 -v' \mu')}{(m_R^2 + M_X^2)^4v'}v_{\rm rel}^2 \right],
\end{align}
where $\lambda_{h S} = (\lambda_{\Phi_1 S} \sin^2 \alpha + \lambda_{\Phi_2 S} \cos^2 \alpha)/4$ is the combination of quartic coupling of $|\Phi_i |^2|S|^2$.  
We then apply these annihilation cross sections to Eq.~(\ref{eq:relic}) to obtain the relic density. 
The spin independent scattering cross section $\sigma_N$ is also given by 
\begin{align}
\sigma_N\approx C\frac{\mu_{DM}^2 (\lambda_{h S} m_n)^2}{4\pi (M_X m_h^2)^2} [{\rm cm}^2],
\label{eq:dd-b}
\end{align}
where  $m_n\approx 0.939$ GeV is the neutron mass, $\mu_{DM}\equiv (1/m_n+1/M_X)^{-1}$, $C\approx(0.287)^2$ is determined by the lattice simulation, and $m_h\approx 125.5$ GeV is the SM-like Higgs.
The latest bound on the spin-independent scattering process was reported by the LUX experiment as an upper limit on the spin-independent (elastic) DM-nucleon cross section, which is approximately $10^{-45}$ cm$^2$ (when $M_X\approx 10^2$ GeV) with the 90 \% confidence level~\cite{Akerib:2013tjd}.\footnote{The sensitivity is recently updated by the same experimental group, which has reached at $2\times10^{-46}$ cm$^2$ at $ {\cal O}$(100) GeV.}
In our numerical analysis below, we set the allowed region for all the mass range of DM to be
\begin{align}
0.11\lesssim \Omega h^2\lesssim 0.13\label{eq:relicexp},\quad \sigma_N\le 10^{-45}{\rm cm^2},
\end{align}
to check the consistency with neutrino mass and muon $g-2$.

\section{Diphoton resonance}
In this section, we discuss the production of heavier CP-even Higgs $H$ and its decays at the LHC 13 TeV.
In our analysis we adopt the alignment limit $\beta - \alpha = \pi/2$ to suppress $H \to W^+W^-/ZZ$ partial decay width and to make lighter CP-even Higgs SM-like as indicated in current Higgs data~\cite{Benbrik:2015evd}.
In our analysis we set mass of $H$ as 750 GeV since it is a well investigated point due to the diphoton excess.
The production of $H$ is given by gluon fusion process via top Yukawa coupling. 
The relevant effective interaction is given by~\cite{Gunion:1989we}
\begin{equation}
{\mathcal L}_{Hgg} = \frac{\alpha_s}{16 \pi} \frac{1}{v \tan \beta} A_{1/2}(\tau_t) H G^a_{\mu \nu} G^{a \mu \nu}, 
\end{equation}
in the alignment limit, where $A_{1/2}(\tau_t) = -\frac{1}{4} [\ln[(1+\sqrt{\tau_t})/(1-\sqrt{\tau_t})] - i \pi ]^2$ with $\tau_t = 4 m_{t}^2/m_H^2$.
Then the production cross section for $m_H = 750$ GeV is $\sigma (gg \to H) \simeq 0.85 \cot^2 \beta$ pb at $\sqrt{s} = 13$ TeV~\cite{Djouadi:2013uqa,Khachatryan:2014wca}.

\begin{figure}[tb]
\begin{center}
\includegraphics[width=70mm]{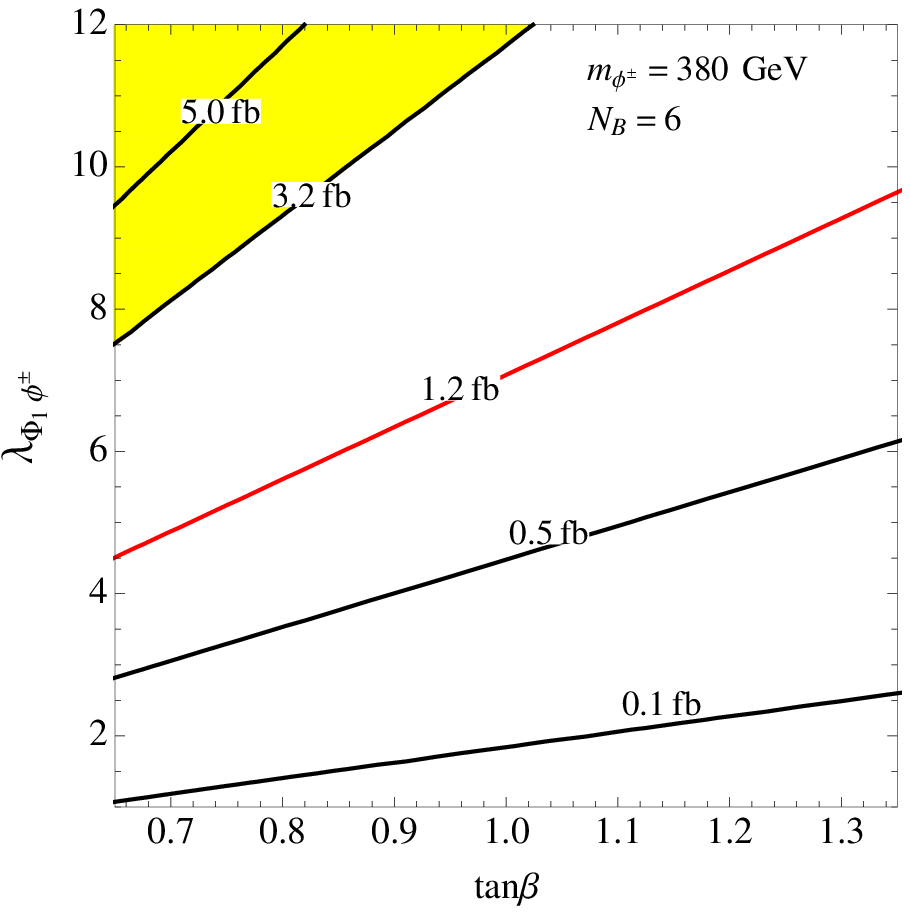}
\includegraphics[width=70mm]{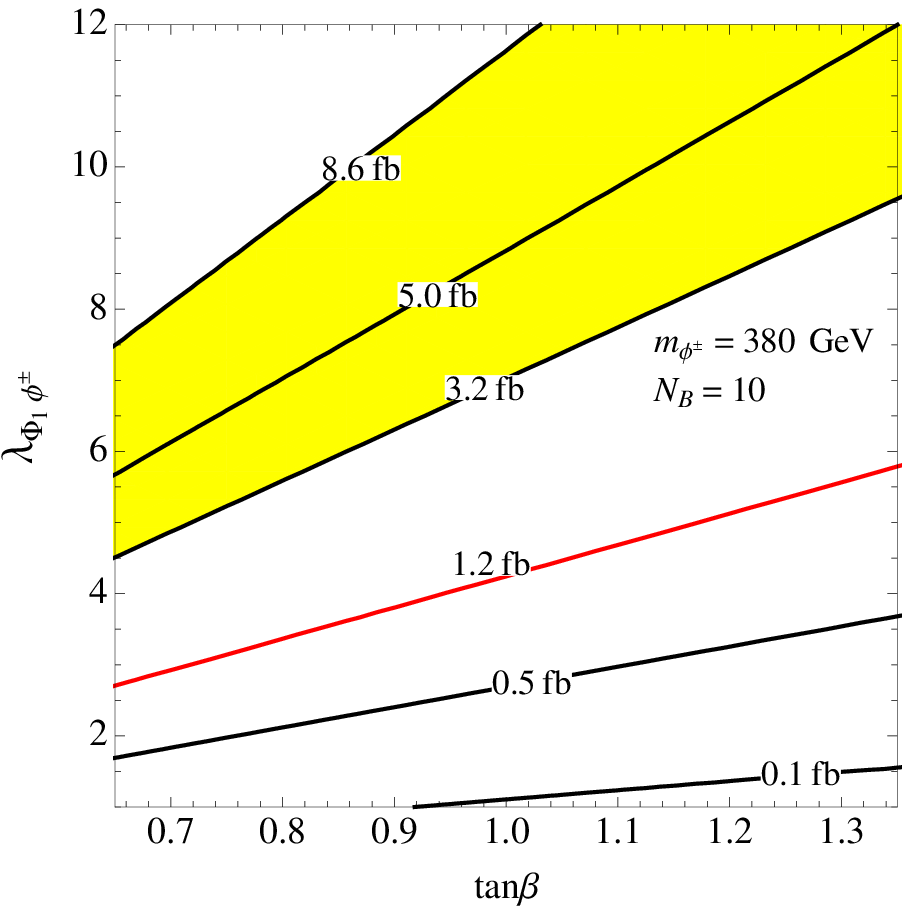}
\caption{The contours of $\sigma (gg \to H) BR(H \to \gamma \gamma)$ (in unit of fb) in $\tan \beta - \lambda_{H \phi^\pm}$ plane.
 The yellow colored region is relevant for explaining the diphoton excess as $3.2$ fb $\lesssim \sigma(pp \to H \to \gamma \gamma) \lesssim 8.6$ fb. 
In addition, the red line shows the upper limit of the cross section indicated by new LHC data in 2016~\cite{ATLAS:2016eeo}.
}   \label{fig:diphoton}
\end{center}\end{figure}

The partial decay width for $H \to t \bar t$ is obtained as
\begin{equation}
\Gamma_{H \to t \bar t} =  \frac{3 m_t^2 \cot^2 \beta}{8 \pi v^2} m_H \sqrt{1 - \frac{4 m_{t}^2}{m_H^2}}.
\end{equation}
The total decay width of $H$ is dominantly given by $t \bar t$ channel as~\cite{hdecay} 
\begin{equation}
\Gamma_{H} \sim \Gamma_{H \to t \bar t} \simeq 32 {\rm GeV} \times \cot^2 \beta.
\end{equation}
We thus note that the width of $H$ tends to large for small $\tan \beta$. 
The $H \to \gamma \gamma$ decay channel is induced through top quark loop and charged scalar loops.
{Here we note that contribution from charged Higgs boson from Higgs doublets is small since coupling constant for $H H^+ H^-$ interaction can not be arbitrary large due to the constraints from precision measurements regarding SM Higgs~\cite{Angelescu:2015uiz}.}
The contribution from exotic fermion is also small since interaction $H \bar e'_L E_R \supset H (s_{\theta_E} c_{\theta_E} \bar E_{1L} E_{1R} -s_{\theta_E} c_{\theta_E} \bar E_{2L} E_{2R} )$ 
provide cancellation between $E_1$ and $E_2$ contributions. 
Thus we focus on the contribution from the loop diagram which contains singlet charged scalars. 
The charged scalar loops are induced by the interactions 
\begin{align}
 V \supset & \sum_{\phi^\pm = k_{1}^\pm, k_{2}^\pm, s^\pm}  (-\lambda_{\Phi_1 \phi^\pm }^{ab} \cos \beta \sin \alpha + \lambda_{\Phi_2 \phi^{\pm}}^{ab} \sin \beta \cos \alpha  ) v h \phi^{a+} \phi^{b-} \nonumber \\
 & + \sum_{\phi^\pm = k_{1}^\pm, k_{2}^\pm, s^\pm}  (\lambda_{\Phi_1 \phi^\pm }^{ab} \cos \beta \cos \alpha + \lambda_{\Phi_2 \phi^\pm}^{ab} \sin \beta \sin \alpha  ) v H \phi^{a+} \phi^{b-},
\end{align}
which are obtained from Eq.~(\ref{eq:potential}).
Since $h \to \gamma \gamma$ branching ratio is consistent with SM prediction, we require $\lambda_{\Phi_1 \phi^\pm }^{ab} \cos \beta \sin \alpha = \lambda_{\Phi_2 \phi^\pm}^{ab} \sin \beta \cos \alpha$ to suppress extra charged scalar contributions. 
Taking into account the alignment limit, we obtain the relevant interactions of $H$ and charged scalars such that 
\begin{equation}
 \sum_{\phi^\pm = k_{1}^\pm, k_{2}^\pm, s^\pm} \lambda_{\Phi_1 \phi^\pm }^{ab}  v \cot \beta H \phi^{a+} \phi^{b-},
\end{equation}
where only diagonal terms contribute to $H \to \gamma \gamma$ process.
The partial decay width is then given by 
\begin{equation}
\Gamma_{H \to \gamma \gamma} = \frac{\alpha^2 m_H^3}{256 \pi^3} \left| \frac{4 \cot \beta}{3 v} A_{1/2}(\tau_t) + \sum_{\phi^{a \pm}} \frac{\lambda_{\Phi_1 \phi^\pm}^{aa} v \cot \beta}{2 m_{\phi^\pm}^2} A_0 (\tau_{\phi^\pm}) \right|^2,
\end{equation}
where $A_0 (x) = -x^2[x^{-1} - [\sin^{-1} (1/\sqrt{x})]^2]$ and $\tau_{\phi^\pm} = 4 m_{\phi^\pm}^2/m_H^2$. 
For simplicity, we apply same value for all $\lambda_{\Phi_1 \phi^\pm}^{aa}$ in calculating the branching ratio.
The masses of the charged scalars are chosen as $380$ GeV($\sim m_H/2$) in order to enhance the value of $A_0 (\tau)$.
In Fig.~\ref{fig:diphoton}, we show the parameter region in $\tan \beta - \lambda_{H \phi^\pm}$ plane which provides products of $H$ production cross section and  branching ratio for diphoton channel.
The yellow colored region show the parameter space which is indicated by the diphoton excess,
$3.2 \, {\rm fb} \leq \sigma (gg \to H) BR(H \to \gamma \gamma) \leq 8.6 \, {\rm fb}$, as a reference
where 1 $\sigma$ error in Refs.~\cite{ATLAS-CONF-2015-081, CMS:2015dxe} is taken into account. 
In addition, the red line shows the upper limit of the cross section indicated by new LHC data in 2016~\cite{ATLAS:2016eeo} such that $\sigma (gg \to H) BR(H \to \gamma \gamma) \leq 1.21$ fb.
Thus we find that sizable cross section, $\sigma (gg \to H) BR(H \to \gamma \gamma)$, can be obtained 
with several sets of charged scalar bosons. 
Thus our model can be tested searching for diphoton signal in future LHC experiments.

\section{ Numerical results}
\begin{figure}[tb]
\begin{center}
\includegraphics[width=80mm]{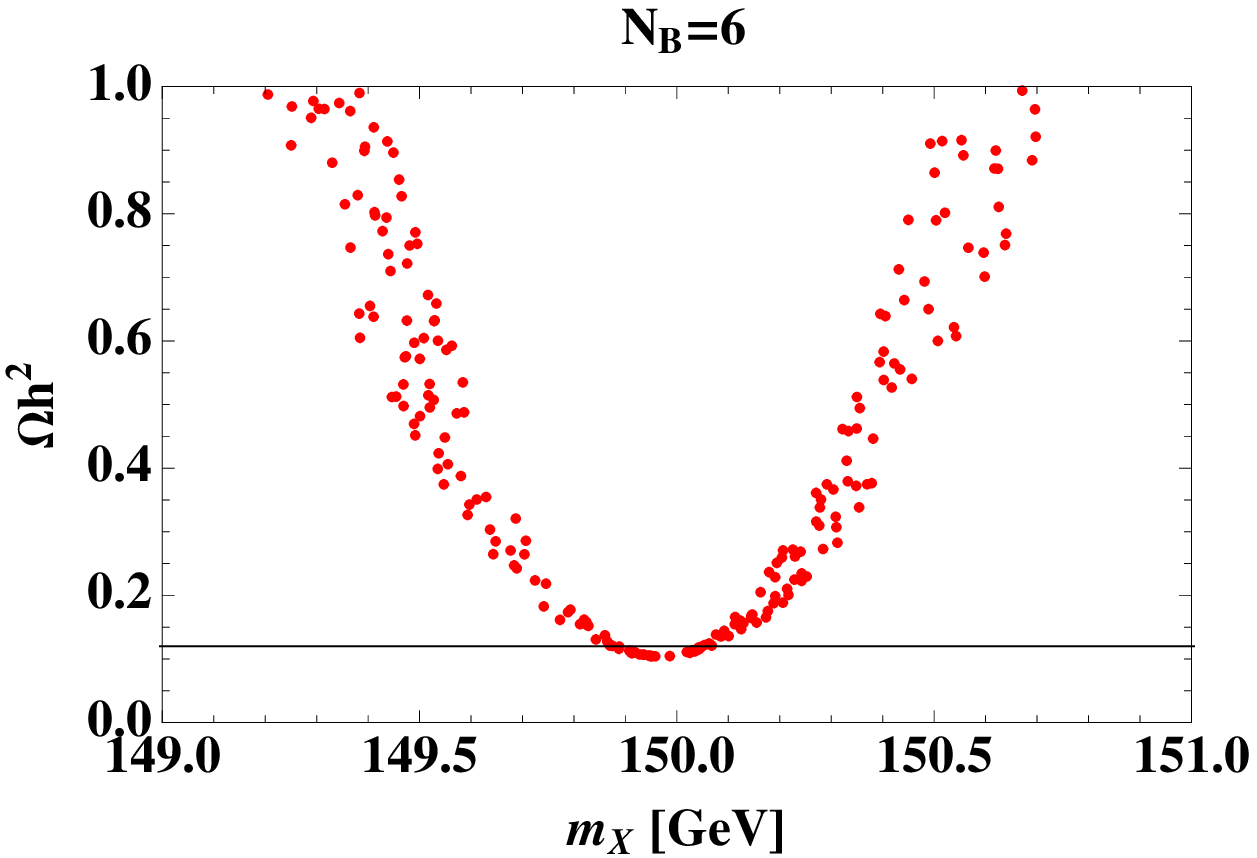}
\includegraphics[width=80mm]{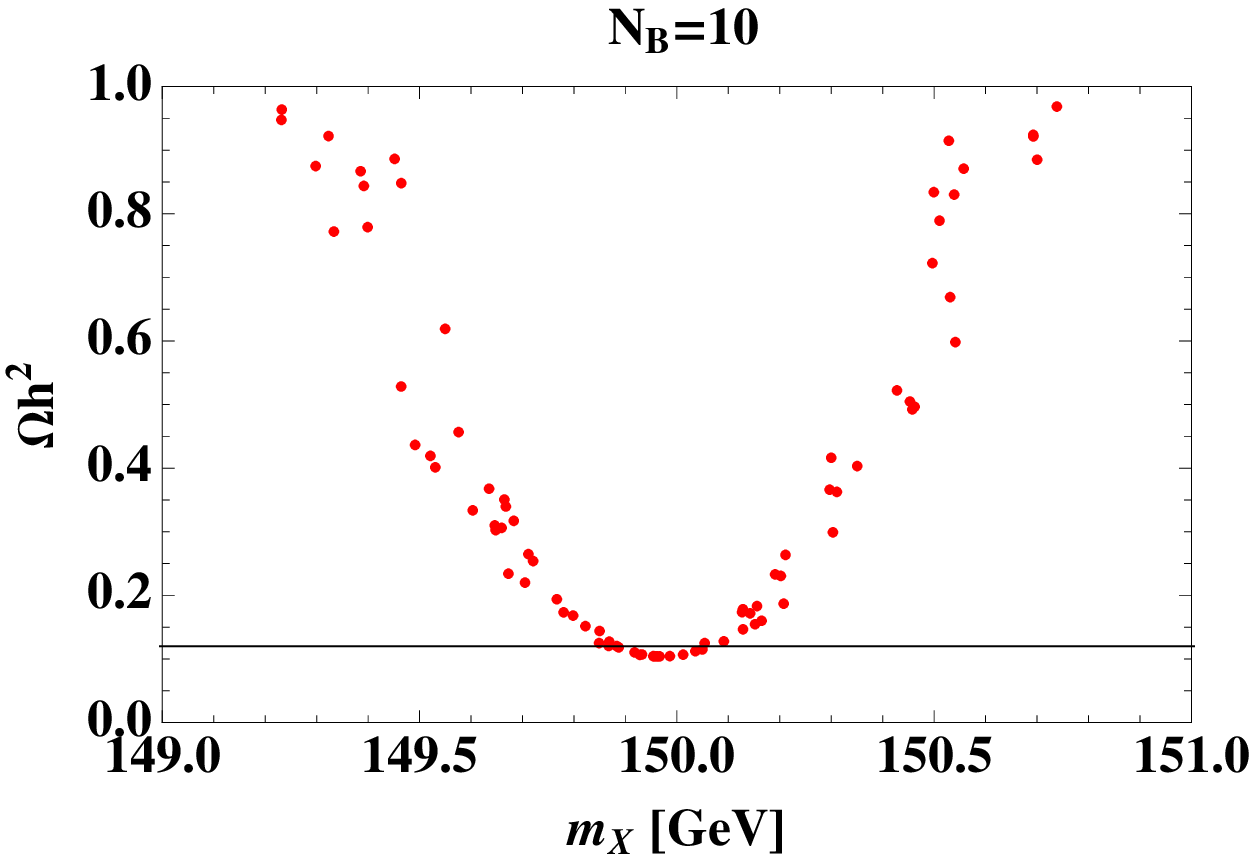}
\caption{{
Our solutions of relic density and the mass of DM within our range in Eq.(\ref{range_scanning}) in case of 
fermion DM that has $10^5$ random sampling points, where 
 left(right) side figure corresponds to $N_B = 6(10)$.
 }
}   \label{fig:scan-FB}
\end{center}\end{figure}
In this section, we perform numerical analysis and show our model can explain neutrino mass, muon $g-2$, relic density of DM and the diphoton excess simultaneously.
Applying the formulas in Sec.~II, we require the neutrino mass and muon $g-2$ to be 
\begin{equation}
0.001\ {\rm eV}\lesssim m_\nu \lesssim 0.1\ {\rm eV}, \qquad 1.0\times 10^{-9}\lesssim \Delta a_\mu\lesssim 4.2\times10^{-9}.
\label{eq:const}
\end{equation}
Firstly, we set masses of all charged singlet scalar as $m_{\phi^\pm} = 380$ GeV, which is required to explain the diphoton excess. 
Now we randomly select values of the {twelve} parameters within the corresponding ranges
\begin{align}
& M_X \in [0.1, 0.18]\ \text{TeV},  \quad \mu=\mu' = v'  \in [0.5,\ 0.6 ]\  \text{TeV}, \nn\\
&[M_{E^{1/2}},\ M_L ,\  M_N, \ m_{R},\ m_{I}] \in [0.5\,\text{}, 0.6\,\text{TeV}], \quad [f,\ y_S] \in [1,\sqrt{4\pi}],  \quad    g \in [0,1], 
\label{range_scanning}
\end{align}
where we universally apply the parameter ranges to the fermion DM and boson DM case  ($M_X$ indicates  $M_{N_R|_{\rm lightest}}$ and $M_{S_I}$), and  fix to be $n=1$, $\lambda_0=10^{-7}$, and $G(x_I)=1$, and $s_{\theta_E}=1/\sqrt2$, $m_\phi=300$ GeV for simplicity and to make it clearer that there exists a resonance solution at the half of the heavier Higgs mass.
Moreover, we set $\lambda_{h S}=0$ to evade the constraint of direct detection search for the boson case in Eq.~(\ref{eq:dd-b}).
Notice here  $k_1^+$ always decays into $X$ (its mass range is $100-180$ GeV), since the charged boson $k_1^+$ (its mass is 380 GeV), couples to the this field that is the lightest field of $N_R(\equiv X)$. 
The heavier $N_R$s can also decay into the final states including the lightest one. 
Then, taking $10^5$ random sampling points,  we show a result for the fermion DM case as can be seen in Fig.~\ref{fig:scan-FB}, in which
our solutions satisfying Eq.~(\ref{eq:const}) are represented on the plane of relic density and the mass of DM within our range in Eq.(\ref{range_scanning}), where left(right) side figure corresponds to $N_B = 6(10)$. One clearly finds that we have a solution at around the half of the heavier Higgs mass $m_\phi/2\approx 150$ GeV for both cases, as can be seen in  Fig.~\ref{fig:scan-FB}.
On the other hand, the boson DM has allowed regions all over the focussed range in Eq.(\ref{range_scanning}), and no specific correlations among their parameters. Thus we abbreviate figures. 

Here we discuss relation among neutrino mass, mass of heavy particle inside the loop diagram Fig.~\ref{fig:neut1}, charged scalar multiplicity $N_B$ and coupling constants. 
For simplicity, exotic charged leptons $E$ and $L'$ are assumed to be heavier than other exotic particles whose masses are taken to be ${\cal O}(1)$ TeV.
Taking other massive parameters $\mu$, $\mu'$ and $v'$ to be also ${\cal O}(1)$ TeV, we obtain order of neutrino mass from Eq.~(\ref{eq:neutrinoM}) such that 
\begin{equation}
m_\nu \sim 3 \times 10^{-7} N_B^6 \lambda_0 (f^2g) \left( \frac{\rm TeV}{M_{E,L}} \right)^4 \ [{\rm GeV}]
\end{equation}
where we took loop factor $G(x_f) \sim 1$ and $\sin 2 \theta_E \sim 1$ for simplicity.
Thus if the coupling constants are ${\cal O}(1)$ we have the upper limit of the heavy particle mass as $M_E \simeq M_L \lesssim \{13, 200, 420\} \ [{\rm TeV}]$ for $N_B = \{1, 6, 10 \}$ 
requiring $0.001\ {\rm eV}\lesssim m_\nu \lesssim 0.1\ {\rm eV}$. 
Therefore high multiplicity of charged scalar allows much heavier scale of masses inside the loop to generate neutrino masses.

\section{ Conclusions and discussions}
We have proposed a new type of radiative neutrino model with global hidden $U(1)$ symmetry, in which neutrino masses are induced at the four loop level, the discrepancy of the muon anomalous magnetic moment to the standard model (SM)  is sizably obtained by using the exotic charged fermions, and both the fermion DM and boson DM candidate can satisfy the observed relic density without conflict of the direct detection searches.
    
 Diphoton resonance has been investigated by adopting two Higgs doublets where the Yukawa coupling with SM fermions is same as Type-II 2HDM. 
Here we have focused on a heavier CP-even neutral boson, which couples to the new charged bosons.
We find that several sets of charged scalar bosons provide a sizable cross section for diphoton signal.
Moreover the constraints from new LHC data in 2016 is also considered which disfavors the previous diphoton excess. 
Although the excess is not confirmed we find that the diphoton resonance search is a good way to test our model which includes many new charged scalar contents.

Finally we have done the numerical analysis satisfying all these physical values or constraints, and shown allowed solutions in terms of the relic density {and the mass of DM} as can be seen in Fig.~\ref{fig:scan-FB}. 
{We have shown a resonance solution at around the half of the heavier Higgs mass $m_\phi/2\approx 150$ GeV for the fermion case,
while the boson DM has allowed regions all over the focussed range in Eq.(\ref{range_scanning}) without specific correlations among their input parameters.} 

{Before closing, we briefly discuss possibility of the new particles production at the LHC. The exotic charged scalar bosons and fermions can be produced via electroweak interactions where they eventually decay into charged lepton and DM candidate due to the interactions shown in Sec.II. Thus one of the signature of our model is the events with charged leptons plus missing transverse energy. The signal events could be observed in LHC-Run2 since we have several new charged particles with O(100) GeV scale mass. The detailed analysis of the signal is beyond the scope of this paper and it will be left as a future study.}


\section*{Acknowledgments}
\vspace{0.5cm}
H.O. thanks to Prof. Shinya Kanemura, Prof. Seong Chan Park, Dr. Kenji Nishiwaki, Dr. Yuta Oriaksa, Dr. Ryoutaro Watanabe, and Dr. Kei Yagyu for fruitful discussions. 
H. O. is sincerely grateful for all the KIAS members, Korean cordial persons, foods, culture, weather, and all the other things.

\end{document}